\documentclass[aip,jcp,preprint]{revtex4-1}

\usepackage[latin1]{inputenc}
\usepackage{graphicx}
\usepackage{dcolumn}
\usepackage{bm}

\begin{document}

\title{Rippled nanocarbons from periodic arrangements of reordered bivacancies in graphene or SWCNTs}

\author{Jean-Marc Leyssale}
\email{leyssale@lcts.u-bordeaux1.fr}
\affiliation{CNRS, Laboratoire des Composites ThermoStructuraux - UMR 5801 Univ. Bordeaux - CNRS - Safran - CEA,F-33600 Pessac}
\author{G\'erard L. Vignoles}
\affiliation{Univ. Bordeaux, Laboratoire des Composites ThermoStructuraux - UMR 5801 Univ. Bordeaux - CNRS - Safran - CEA,F-33600 Pessac}
\author{Antoine Villesuzanne}
\affiliation{CNRS, Univ. Bordeaux, ICMCB, 87 Av. Dr. A. Schweitzer, 33608 Pessac Cedex, France}

\date{\today}

\begin{abstract}
We report on various nanocarbons formed from a unique structural pattern containing two pentagons, three hexagons and two heptagons, 
resulting from local rearrangements around a divacancy in pristine graphene or nanotubes.
This defect can be inserted in sheets or tubes either individually or as extended defect lines. Sheets or tubes containing only this 
defect as a pattern can also be obtained.
These fully defective sheets, and most of the tubes, present a very pronounced rippled (wavy) structure and their energies are lower than other structures 
based on pentagons and heptagons published so far. 
Another particularity of these rippled carbon sheets is their ability to fold themselves into a two-dimensional porous network of interconnected
tubes upon heat treatment as shown by hybrid Monte Carlo simulations. Finally, contrary to the common belief that pentagon/heptagon based structures are metallic, this work 
shows that this defect pattern should give rise to semi-metallic conduction.
\end{abstract}

\maketitle

\section{Introduction}
\label{intro}

Graphene may contain defects. This is now an established fact as it has been directly imaged, thanks to atomic resolution electron microscopy 
\cite{hashimoto_nat_04,meyer_nl_08,lahiri_natnt_10,kotakoski_prl_11}. Defects in graphene have first been obtained 
by reconstruction around vacancies formed during electron irradiation and leading to pseudo amorphous domains made of pentagons, hexagons and 
heptagons \cite{hashimoto_nat_04,meyer_nl_08,kotakoski_prl_11}. They can also show some ordering as observed by Lahiri {\it et al.} \cite{lahiri_natnt_10}, 
who describe an extended defect line, made of a periodic repetition of two adjacent pentagons and an octagon, in a graphene sheet epitaxially 
grown on a Ni substrate with a tilt boundary. 

Defects affect the properties of graphene, for the bad in most cases but also sometimes for the good as in 
the case of Lahiri {\it et al.} where the line defect gives to the sheet the electronic properties of a one-dimensional metallic wire \cite{lahiri_natnt_10,
gunlycke_prl_11,recher_phys_11}.
The quest for stable space-filling one- or two-dimensional structural patterns has started long ago in the virtual world, actually much before the experimental 
characterization of such defects, thanks to {\it ab-initio} and force-fields based modeling approaches \cite{crespi_prb_96,terrones_prl_00,lusk_prl_08,lusk_car_09}. Somehow, the 
computer-aided design of new graphene allotropes with targeted electronic or mechanical properties is even becoming a new area in carbon science \cite{carr_natnt_10}.
Among all the structures proposed so far, hexagonal haeckelite $H_{5,6,7}$ has the lowest energy above graphene \cite{terrones_prl_00}. This can be explained by the absence
of adjacent pentagonal rings and its high density of pentagon-heptagon ($C_5/C_7$) pairs, a remarkably stable pattern \cite{grantab_sci_10,jeong_prb_08,yazyev_prb_10,
albertazzi_pccp_99}.

Two years ago, some of us have developed a new computational method based on a combination of image processing and atomistic simulation, the image guided 
atomistic reconstruction (IGAR) method \cite{leyssale_apl_09}, aiming at building atomistic models of dense, disordered yet nanostructured, carbons on the basis of 
their high resolution transmission electron microscopy (HRTEM) lattice fringe images. In this approach, a 3D analogue of the 2D HRTEM image is constructed and serves as an external
potential field, pushing the atoms towards the dark areas of the image during a liquid quench molecular dynamics simulation of a carbon system, in which carbon-carbon interactions 
are taken into account by a reactive empirical bond order potential \cite{brenner_jpcm_02,stuart_jcp_00}. 

\begin{figure}[htbp]
\begin{center}
\includegraphics*[width= 12 cm]{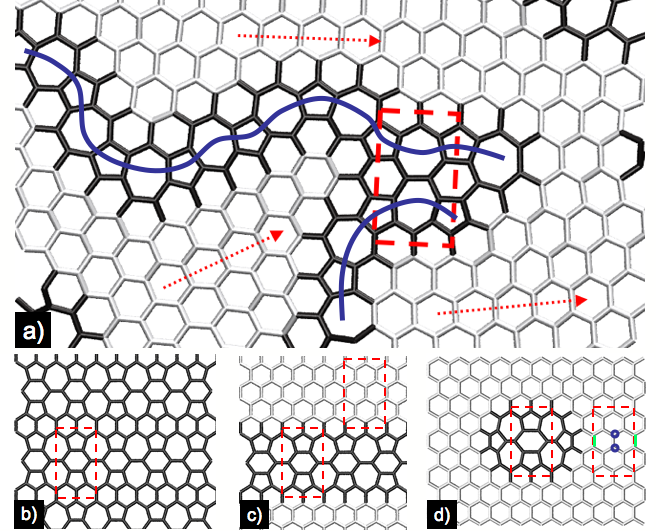}
\end{center}
\caption{Snapshot of a carbon sheet taken from an atomistic model of a pyrolytic carbon, reconstructed using the IGAR method\cite{leyssale_apl_09} (a). 
Bonds between threefold atoms belonging only to hexagons are shown in light grey, other bonds in dark grey; two (blue) lines show the dislocations, 
made of a succession of pentagon/heptagon pairs, between disoriented (as indicated by the (red) dotted arrows) hexagonal domains. The dashed red 
rectangle shows a space-filling structural pattern containing two pentagons, three hexagons (including a 90$^\circ$ rotated one) and two heptagons. 
This pattern can be used to form a fully defective carbon sheet (b), a line (c) or a point (d) defect in a graphene sheet. Dashed rectangles on panels 
(c) and (d) show the geometrical equivalence between the defect unit cell and the one of graphene. It is obtained from graphene by creating a divacancy 
(blue circles on panel (d)) and by a vertical to horizontal flip of the first neighbor parallel C-C bonds (green sticks on panel d).}
\label{defects}
\end{figure}

A snapshot of a carbon sheet taken from a model of a rough laminar pyrolytic carbon \cite{bourrat_car_06}, reconstructed with this method, is shown Figure \ref{defects}a. 
As can be seen, the sheet is made of disoriented purely hexagonal domains (light grey) bound together by defective planar areas made essentially of 
$C_5/C_7$ pairs (dark grey) and more precisely lines of $C_5/C_7$ pairs as indicated by the blue solid lines. 
This agrees very well with some recent theoretical studies on dislocations in graphene showing the high stability of the $C_5/C_7$ pattern 
\cite{grantab_sci_10, yazyev_prb_10,jeong_prb_08}. Also, and even though the way this model was obtained has nothing in common with the irradiation
 of a single graphene layer, its structure is amazingly close from those reported by Meyer and co-workers \cite{meyer_nl_08,kotakoski_prl_11}. 
Looking at this figure, the structural pattern highlighted by the dashed rectangle appeared to be of a particular interest to us. Indeed, this domain, 
made of two pentagons, three hexagons (including a 90$^\circ$ rotated one) and two heptagons, allows for the construction of an entire carbon sheet based 
on this defect (see \ref{defects} b), noted $C_5^2C_6^3C_7^2$ hereafter, as well as for the insertion of a line (\ref{defects} c) or a point 
(\ref{defects} d) defect in graphene. Note that the latter point defect has been very recently observed in a HRTEM \cite{kotakoski_prl_11}.

In what follows, we report on the structure and properties of different nano-sheets and tubes based on this pattern as obtained from density functional theory (DFT) 
and force field calculations. Methodological details are given in section \ref{methods}, results are presented and discussed in section \ref{results}.

\section{Computational methods} \label{methods}

\subsection{Density functional theory calculations}

Geometry optimization of the $C_5^2C_6^3C_7^2$ unit cell has been achieved using periodic density functional theory calculations.
The VASP package \cite{kresse_cms_96, kresse_prb_93, kresse_prb_96} was used with projector augmented wave pseudopotentials
\cite{blochl_prb_94, kresse_prb_99} to describe the valence-ion core interactions, and the Perdew-Wang-Ernzerhof generalized
gradient approximation for the exchange-correlation potential \cite{perdew_prl_96}.
Atomic positions and lattice parameters were relaxed, and energy criteria of $10^{-4}$ and $10^{-3}$ eV/cell were used for electronic
and structural convergences, respectively. A 15 \AA \ interlayer distance was kept fixed for all calculations, and up to 450-point 2D grids
were used for Brillouin zone sampling. The kinetic energy cutoff for plane-wave expansions was 400 eV.

\subsection{Hybrid Monte Carlo simulations}

Geometry optimizations of large $C_5^2C_6^3C_7^2$ planes and tubes were achieved using hybrid Monte Carlo (HMC) simulations \cite{mehlig_prb_92}
coupled to a simulated annealing approach. The interaction energy between carbon atoms was described by the reactive AIREBO potential \cite{stuart_jcp_00} after
removal of the van der Waals terms of this potential. Sheets and tubes were placed in an orthorhombic simulation cell with periodic boundary
conditions in all directions.
Sheets (respectively tubes) were placed on the $xy$ plane (resp. along the $z$ axis) of the box and a large dimension (500 \AA) was given to the
$z$ dimension (resp. $x$ and $y$ dimensions) to ensure the simulation of isolated infinite systems.

\noindent Two kinds of configuration updates were used (chosen at random) in our HMC implementation, allowing for a proper sampling of configurational space in the
isothermal-isobaric ($NPT$) ensemble:

\begin{itemize}
\item Short constant energy molecular dynamics (MD) trajectories with initial velocities drawn from a Maxwell distribution at the suited temperature accepted with
the usual canonical HMC acceptance criterion \cite{mehlig_prb_92}.
\item Random contraction or dilation of the volume in one dimension of the system ($x$ or $y$ for sheets, $z$ for tubes) accepted or rejected with the standard
isothermal-isobaric Monte-Carlo acceptance criterion \cite{allen_tildesley}
\end{itemize}

\noindent Simulations were initiated from the fully planar sheets or fully cylindrical tubes with cell dimensions taken from the corresponding graphene sheets or
pristine tubes. Constant $NPT$ HMC simulations of $10^4$ steps were then performed at zero pressure and a given temperature (note that the position of one carbon
atom with respect to the simulation cell was held fixed in order to avoid global translation or rotation of the system). After that, the temperature was scaled by
a factor 0.9 and the process repeated until convergence of both potential energy and geometry. In the HMC simulations, the probabilities for MD and volume
move attempts were respectively of 70 \% and 30 \%. The length of the MD trajectories (both in terms of the number of steps and of the timestep length) as well as
the magnitude of the maximum volume change attempt were periodically adjusted to yield acceptance ratios of around 50 \% for both kinds of moves. We used an initial
temperature of 500 K and stopped the simulated annealing when the temperature was below $10^{-2} \; K$, which appeared to be sufficient for a full convergence of
geometry and energy.

In the last part of the paper (see Figure 5) HMC was not used anymore to cool a system down to 0K but to heat it up, as slowly as possible, to high temperatures.
In that case similar simulations were used except that (i) the initial configuration was taken from the previously minimized system at 0 K; that (ii)
$5 \times 10^4$ HMC steps were performed at each temperature and that (iii) the target temperature was slowly increased by steps of 10 K.

\section{Results} \label{results}

\begin{figure}[htbp]
\begin{center}
\includegraphics*[width= 12 cm]{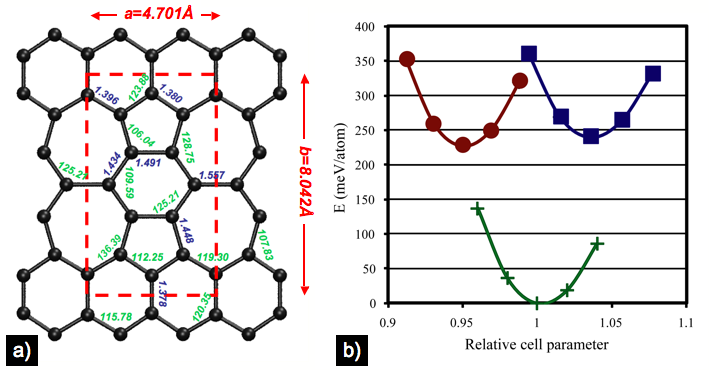}
\end{center}
\caption{Optimized unit cell (a) and relative stability and potential well of the $C_5^2C_6^3C_7^2$ sheet as compared to graphene and hexagonal haeckelite (b) as computed from
periodic DFT calculations. 
Red, blue and green labels on panel (a) respectively indicate unit cell parameters (in \AA), bond lengths (in \AA) and angles (in degrees); 
red circles, blue squares and green crosses on panel (b) give the energy as a function of the relative cell parameter for $C_5^2C_6^3C_7^2$, haeckelite and graphene 
(taken as the reference for energy and cell dimension) sheets respectively.}
\label{LeyOPT_UC}
\end{figure}

Figure \ref{LeyOPT_UC}a shows the unit cell of the $C_5^2C_6^3C_7^2$ as optimized by periodic density functional theory (DFT) calculations. 
No significant deviation from the rectangular lattice and no departure from planarity were found 
in our DFT calculations; however, this holds for the 14-atoms unit cell of Figure \ref{LeyOPT_UC}a and does not exclude non-planar configurations in supercells, {\it i.e.} with 
larger degrees of freedom, as we will see in what follows. The energy of the optimized structure (see Figure \ref{LeyOPT_UC}b) is only of 227 meV/atom above graphene, making 
it the lowest energy defect-containing carbon sheet reported so far. To our knowledge, it was formerly the case of hexagonal haeckelite, with an energy of 240 meV/atom 
above graphene (note that we compare to those structures as optimized using the same methodology that we applied to the $C_5^2C_6^3C_7^2$ sheet). 
Also, Figure \ref{LeyOPT_UC}b shows that the potential wells of the three structures are very similar. 

\begin{figure}[htbp]
\begin{center}
\includegraphics*[width= 11 cm]{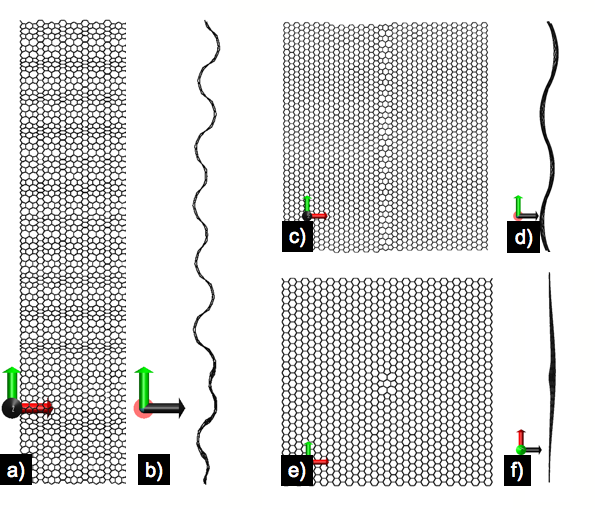}
\end{center}
\caption{Snapshots of a periodic $5 \times 21$ $C_5^2C_6^3C_7^2$ sheet (a,b), a $1 \times 21$ line defect (c,d) and a single point defect (e,f) inserted in graphene sheets. Geometries have been obtimized using HMC simulated annealing and the AIREBO force field. (a,c,e): plane view; (b,d,f): lateral view.}
\label{sheets}
\end{figure}

We present now the properties of large carbon sheets and nanotubes based on the $C_5^2C_6^3C_7^2$ pattern as obtained from cell and geometry optimizations using 
hybrid Monte Carlo \cite{mehlig_prb_92} (HMC) simulated annealing and the AIREBO force field \cite{stuart_jcp_00}.
Figure \ref{sheets} shows a front (a) and a side (b) view of an optimized periodic $C_5^2C_6^3C_7^2$ sheet made of $5 \times 21$ unit cells. 
The energy of this sheet, as given by the AIREBO potential is of $-7.131$ eV/atom, $279$ meV/atom above graphene as computed with the same potential ($-7.408$ eV/atom). 
Relative energies of the $C_5^2C_6^3C_7^2$, haeckelite and graphene sheets given by this potential are in fair qualitative agreement with DFT calculations (haeckelite is 
found at $323$ meV/atom above graphene with this potential). 
However, a major difference is observed between haeckelite and $C_5^2C_6^3C_7^2$ sheets. Indeed, if the geometry optimization of a large haeckelite sheet reveals an almost 
planar geometry, Figures \ref{sheets}a and \ref{sheets}b unambigously unravel a wavy structure developing itself along the $a$ unit cell direction, the direction of the 
adjacent hexagons lines in the structure. Optimizing the number of unit cells in that direction we found out that the best possible periodicity is four unit cells along the 
$a$ axis, giving rise to an optimal energy of $-7.135$ eV/atom. 
The line defect of Figures \ref{sheets}c and \ref{sheets}d is also of particular interest. First, and even though it is not easy to directly compare the energies of different 
line defects, it should be of rather low energy as it is made of a remarkably stable pattern (the $C_5^2C_6^3C_7^2$). Second, while other observed \cite{lahiri_natnt_10} or 
predicted \cite{lusk_prl_08,gunlycke_prl_11} line defects usually induce a curvature of the sheet perpendicularly to the line, Figure \ref{sheets}d clearly shows that the 
$C_5^2C_6^3C_7^2$ pattern gives a wavy structure to the sheet, along the line defect, in close similarity with the wavy structure developed on the perfect $C_5^2C_6^3C_7^2$ 
sheet, though attenuated by the large hexagonal domains on both sides. 
The point defect of Figures \ref{sheets}e and \ref{sheets}f also has an excellent energetic stability, but affects only very locally the structure of the sheet. 
It is an antisymmetric defect with an inversion center localized on the center of the inverted hexagon. We recall that this point defect has been very recently imaged by 
Kotakoski {\it et al.} during a HRTEM observation of an irradiated graphene layer\cite{kotakoski_prl_11}.

\begin{figure}[htbp]
\begin{center}
\includegraphics*[width= 11 cm]{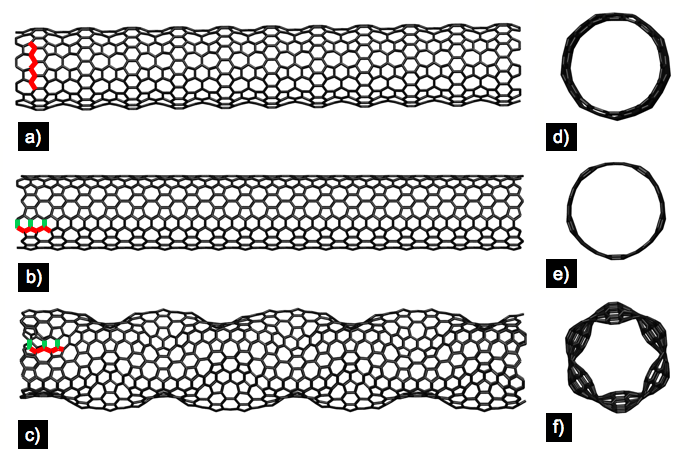}
\end{center}
\caption{Snapshots of carbon nanotubes built from the $C_5^2C_6^3C_7^2$. Geometries have been obtimized using HMC simulated annealing and the AIREBO force field. 
In analogy with pristine nanotubes, tubes having the lines of hexagons along (b,c) and perpendicular to (a) the tube 
axis are respectively noted as zigzag (ZZ) and armchair (AC). a: side view of a zigzag tube made by folding a ribbon of nine unit cells along the $a$ axis (ZZ9); b: side view of an 
armchair tube made by folding a ribbon of five unit cells along the $b$ axis (AC5); c: side view of an armchair tube made by folding a ribbon of six unit cells along the $b$ axis (AC6). 
(d,e,f): top views of tubes (a,b,c).} 
\label{tubes}
\end{figure}

\begin{figure}[htbp]
\begin{center}
\includegraphics*[width= 11 cm]{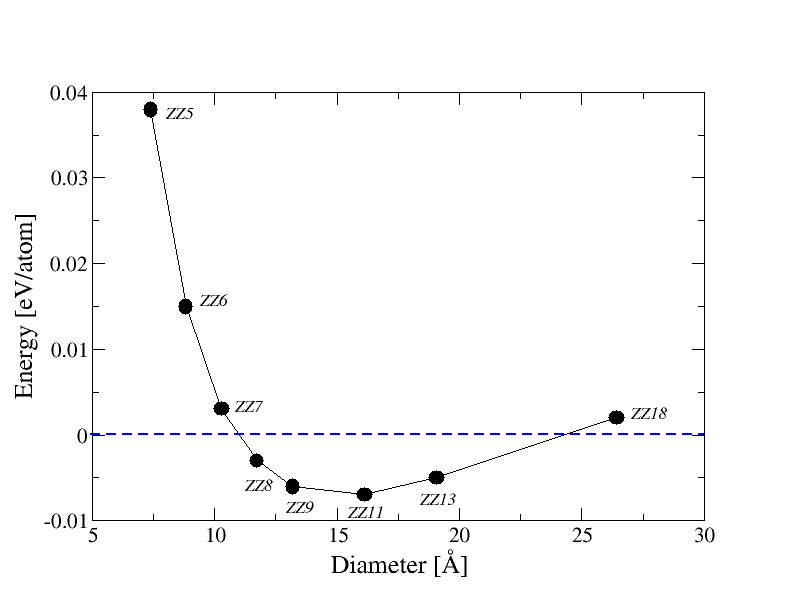}
\end{center}
\caption{Energy of zigzag tubes as a function of their diameter as obtained from hybrid Monte Carlo optimizations (the origin of energy is the lowest energy $C_5^2C_6^3C_7^2$ sheet).}
\label{tubeZZ}
\end{figure}

\begin{figure}[htbp]
\begin{center}
\includegraphics*[width= 11 cm]{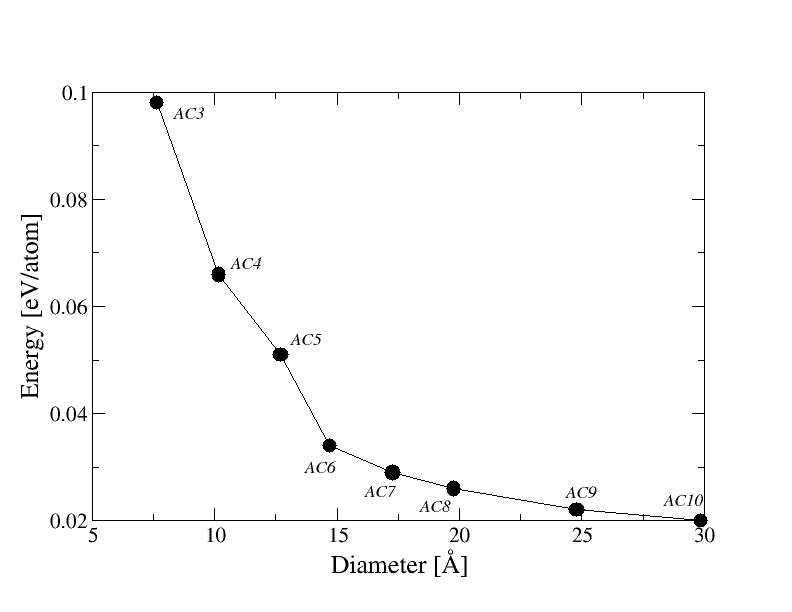}
\end{center}
\caption{Same as Figure \ref{tubeZZ} for AC tubes.}
\label{tubeAC}
\end{figure}

As already said, the $C_5^2C_6^3C_7^2$ pattern likes curvature. This should then be a particurlarly stable pattern to build nanotubes with. 
In close similarity with pristine, graphene based zigzag (ZZ) and armchair (AC) tubes we have built $C_5^2C_6^3C_7^2$ \ ZZ and AC tubes by folding $C_5^2C_6^3C_7^2$ ribbons along 
respectively the $a$ and $b$ axes of the pattern's unit cell (see also the caption of Figure \ref{tubes}). 
We present now some results of HMC geometry optimizations on ZZ and AC tubes of different diameters, ranging from $7$ to $30$ \AA, and with lengths around 9 nm.
Figure \ref{tubes} present snapshots of the optimized ZZ9, AC5 and AC6 tubes. Their diameters, taken as the double of the mean atomic distance from the tube axis, are respectively $13.2$, 
$12.7$ and $14.7$ \AA. All these tubes have rather low energies, respectively $-7.141$, $-7.084$ and $-7.101$ eV/atom for ZZ9, AC5 and AC6; and present different morphologies.
All zigzag tubes studied in this work have a "necklace of pearls" shape as shown on Figures \ref{tubes}a and \ref{tubes}d for ZZ9, in close similarity with some haeckelite tubes
\cite{biro_car_04}. Low diameter armchair tubes (from AC3 to AC5) have a fairly cylindrical morphology (see the snapshots of tube AC5 on Figs \ref{tubes}b and \ref{tubes}e), 
typical of graphene based tubes. However, from AC6 (14.7 \AA) to AC12 (29.9 \AA), armchair tubes develop a pronouced helical morphology (see the snapshots of tube AC6 tube 
on Figs \ref{tubes}c and \ref{tubes}f).

\begin{figure}[htbp]
\begin{center}
\includegraphics*[width= 11 cm]{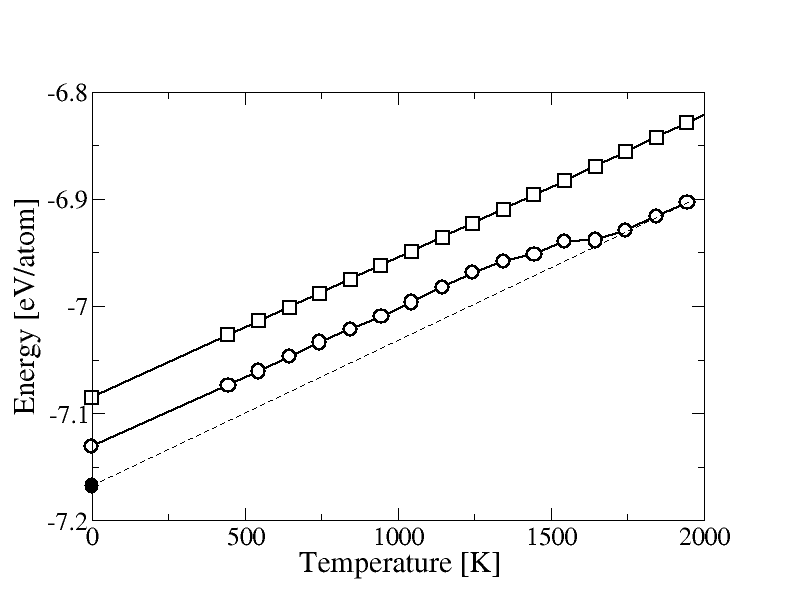}
\end{center}
\caption{Evolution of the energy with respect to temperature during a HMC simulation of the heat treatment of a periodic $17 \times 9$ $C_5^2C_6^3C_7^2$ sheet from 0 to 2000 K 
(empty circles) and cooling it back to 0 K (filled circle). The same simulation as applied to a haeckelite sheet of similar size is indicated shown for comparison (empty squares).} 
\label{warmup}
\end{figure}

On Figure \ref{tubeZZ} we plot the energy of ZZ tubes as a function of their diameter relatively to the sheet of lowest energy (-7.135 eV/atom). As expected, the energy per atom, 
around 40 meV above the $C_5^2C_6^3C_7^2$ sheet for the smallest tube, diminishes with the tube diameter at low diameters thanks to the relaxation of curvature stresses. However,
the behaviour of the curve above 10 \AA \ is particularly unusual. First, we can see that ZZ tubes with diameters in the range [12:25 \AA], from ZZ8 to ZZ13 here, have lower
energies than the rippled sheet. Second, the curve goes through a minimum around 15 \AA, with the minimum energy tube found here, ZZ11, being 7 meV/atom lower than the sheet. Finally, 
above 25 \AA \ diameter, ZZ18 tube has a positive energy with respect to the sheet. These results indicate once again the high affinity of the $C_5^2C_6^3C_7^2$ pattern for curvature 
along the lines of hexagons ($a$ axis) and tend to show that there exists an optimal curvature for this pattern which corresponds roughly to a 15 \AA \ diameter ZZ tube.

\begin{figure}[htbp]
\begin{center}
\includegraphics*[width= 11 cm]{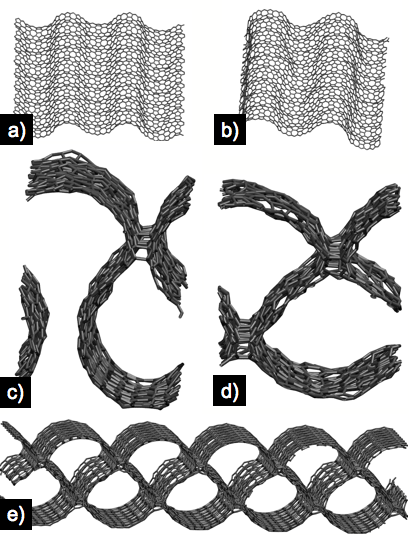}
\end{center}
\caption{
Snapshot of the $C_5^2C_6^3C_7^2$ sheet of Figure \ref{warmup} after annealing at 500 K (a), 1000 K (b), 1500 K (c) and 2000 K (d). A snapshot of the system cooled back to 0 K 
(filled circle on Figure \ref{warmup}) is also shown (e). (note that five periodic replicas of the system, along the horizontal direction, are shown on (e) to highlight the 2D 
nature of the material and its network of aligned (1D) porous channels).}
\label{warmup_snaps}
\end{figure}

Figure \ref{tubeAC} is equivalent to Figure \ref{tubeZZ} for armchair tubes. The first points to note are that AC tubes present higher energies than ZZ tubes at equivalent diameters
and that all the AC tubes studied in this work have higher energies than the lowest rippled sheet. For instance, the most stable AC tube studied in this work, AC10 ($\phi = $ 30 \AA),
is found 20 meV/atom above the sheet. Unlike ZZ tubes, the evolution of AC tubes energy with respect to diameter is monotonically decreasing; however, a step is observed between the low diameter cylindrical tubes (AC3 to AC5) and the larger helical tubes (AC6 to AC10).

\begin{figure}[htbp]
\begin{center}
\includegraphics*[width= 11 cm]{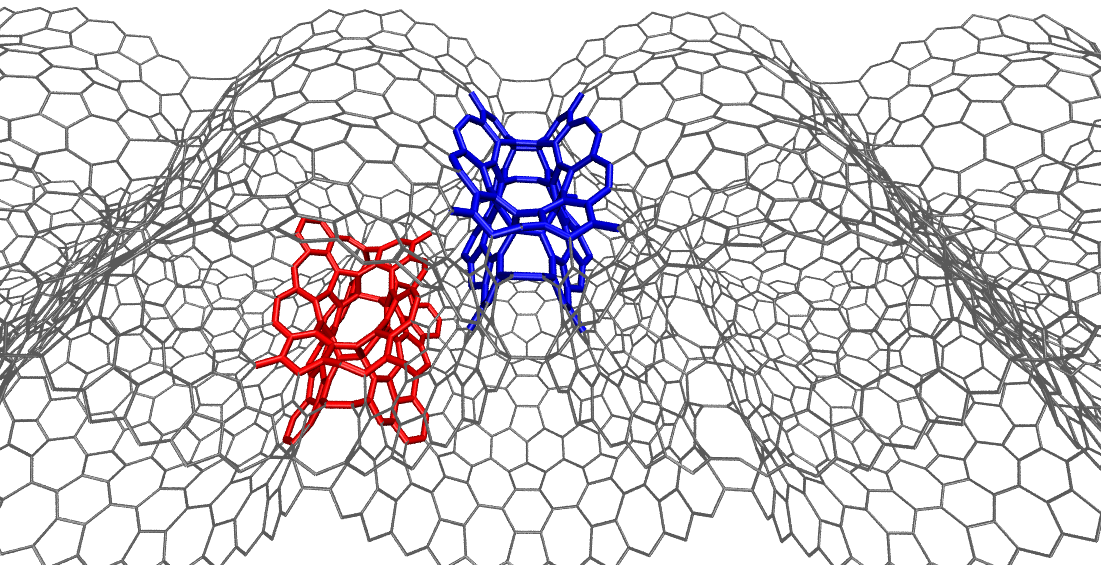}
\end{center}
\caption{Enlargement of Figure \ref{warmup_snaps}f with two coloured regions highlighting the open windows to an upper tube (red) and a lower tube (blue).}
\label{warmup_pores}
\end{figure}

In order to investigate the thermal stability of the $C_5^2C_6^3C_7^2$ sheet we now present results of a HMC simulation in which a periodic sheet of 2142 atoms 
was slowly heated from 0 to 2000 K. The evolution of the energy per atom as a function of temperature is shown on Figure \ref{warmup} and some snapshots of the system
during the heating process are given Figure \ref{warmup_snaps}. As can be seen on Figure \ref{warmup}, the energy evolves rather linearly with temperature at low temperatures and 
the sheet shows three well defined waves (see Figure \ref{warmup_snaps}a). Around 1000 K, a slight drop
in the energy increase is observed and the sheet evolves from three low amplitude waves (Figure \ref{warmup_snaps}a) to two well-pronounced waves (Figure \ref{warmup_snaps}b). 
The energy then keeps on increasing linearly until another drop at around 1400 K. At that point, two parts of the sheet enter into contact and form a regular bridge, 
containing $sp^3$ hybridized atoms, all along the $b$ unit cell axis (Figure \ref{warmup_snaps}c). 
This is shortly followed by the formation of a second, and symmetric, bridge around 1700 K to form an almost perfect periodic two-dimensional network of interconnected 
tubes (Figure \ref{warmup_snaps}d). 
This rippled nanocarbon obtained at 2000 K, shown on Figure \ref{warmup_snaps}d, has then been optimized by a HMC simulated annealing quench down to 0 K, the latter structure being 
shown Figure \ref{warmup_snaps}e. 
Its energy, of $-7.167$ eV/atom, is $32$ meV/atom below the one of the initial sheet, indicating, once again, the affinity of the $C_5^2C_6^3C_7^2$ pattern for high curvatures. 
The diameters of the interconnected tubes forming this porous structure are actually very similar to those of the ZZ9 tube presented on Figures \ref{tubes}a and \ref{tubes}d, 
and correspond to the optimum curvature for zigzag tubes (see Figure \ref{tubeZZ}).
Looking at Figure \ref{warmup_snaps}e, the structure of this carbon immediately reminds of some shock protection materials used in everydays life packaging and there is no major 
risk taken in guessing that its mechanical properties would be particularly unusual. Also, it should be regarded as a very interesting adsorbent due to its aligned network of 
high aspect ratio tubular pores.

Figure \ref{warmup_pores} shows an enlargement of Figure \ref{warmup_snaps}d, to unravel the porous nature and accessibility of this new form of carbon. As can be seen on this 
figure, those pores are extremely easy to access, thanks to numerous windows all along the structure, probably facilitating any adsorption/desorption process
with respect to other CNT based adsorbents.

Finally, it is interesting to note, as illustrated on the top of Figure \ref{warmup}, that the same heating simulation on haeckelite shows a perfectly linear evolution of energy 
with temperature which corresponds to a perfect conservation of the sheet structure (not shown), aside from thermal vibrations.

\begin{figure}[htbp]
\begin{center}
\includegraphics*[width= 11 cm]{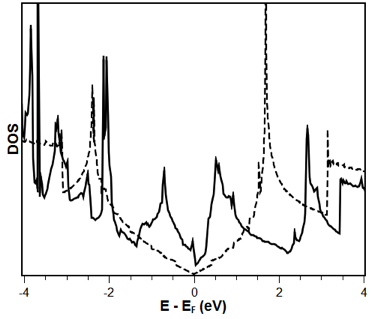}
\end{center}
\caption{Electronic density of states (DOS) of the $C_5^2C_6^3C_7^2$ (straight line) and graphene (dashed line) unit cells as computed from DFT calculations.}
\label{dos}
\end{figure}

\section{Conclusion}

We have proposed new rippled carbon planes and tubes based on a new structural pattern, the $C_5^2C_6^3C_7^2$ pattern, recently observed in some atomistic models 
of pyrocarbons. These structures can be either entirely made of this pattern or inserted as line or point defects in graphene or graphene-based tubes. 
Both DFT and force field energy minimisations show that structures built from this pattern have low energies (actually, the lowest energies reported so far) for structures with a 
high density of non-hexagonal rings.
Another evidence of the high stability associated to this pattern is that the point defect of Figures \ref{sheets}e and \ref{sheets}f has been recently imaged in a TEM \cite{kotakoski_prl_11}.
Structures based on the $C_5^2C_6^3C_7^2$ pattern love curvature and show rather unusual properties: (i) sheets have a pronounced wavy structure; (ii) line defects develop a parallel wavy 
structure; (iii) tubes may show lower energies than sheets; and (iv), sheets are able to close themselves under heat treatment to form a 2D ordered nanoporous carbon. 
Finally, the highest surprise stemming from this defect type maybe lies in its electronic properties. Indeed, it is now a common thought that pentagon and heptagon containing carbon 
structures are metallic \cite{crespi_prb_96,terrones_prl_00,rocquefelte_nl_04, gunlycke_prl_11,lahiri_natnt_10,kahaly_sma_08}. 
As can be seen on the density of states plot of Figure \ref{dos}, and despite its high density of pentagons and heptagons, the $C_5^2C_6^3C_7^2$ network is semi-metallic, alike 
graphene (although its DOS has more features than the one of graphene due to its lower symmetry and the more complex connectivity path of its carbon network).

\section*{Acknowledgements}

Access to the computer ressources of the Mesocentre de Calcul Intensif en Aquitaine (MCIA) is gratefully acknowledged.



%

\end{document}